
%
\magnification=1200
\hsize=6 truein
\def\xb{{\bf x}}\def\kb{{\bf k}}\def\ck{c_{\kb}}
\def\o{\omega}\def\g{\gamma}
\def\lc{\lambda_{cut}}
\def\D{\Delta}
\def\one{{\it 1}}\def\half{{\it {1\over2}}}
\def\xbo{\xb_{\o}}
\def\matilde{\mathaccent "7E}\def\dot2{\mathaccent "7F}
\null
\hfill {KFKI-RMKI-23-APRIL-1993}
\vskip 4cm
\centerline{\bf Calculation of X-Ray Signals}
\centerline{\bf from K\'arolyh\'azy Hazy Space-Time}
\vskip 1cm
\centerline{L. Di\'osi\footnote{$^1$}{diosi@rmki.kfki.hu}
       and B. Luk\'acs\footnote{$^2$}{lukacs@rmki.kfki.hu}}
\centerline{KFKI Research Institute for Particle and Nuclear Physics}
\centerline{H-1525 Budapest 114, POB 49, Hungary}
\vskip 4cm
\line{\bf Abstract\hfill}

K\'arolyh\'azy's hazy space-time model, invented for breaking down
macroscopic interferences, employs wave-like gravity disturbances.
If so, then electric charges would radiate permanently. Here we discuss the
observational consequences of the radiation. We find that such
radiation is excluded by common experimental situations.
\vfill
\eject
\baselineskip=18pt plus 3pt minus 1pt
\line{\bf 1. Introduction\hfill}

In a series of papers [1-3], K\'arolyh\'azy {\it et al.} discussed
the idea that space-time haziness puts an eventual limit on
quantum coherence of massive systems.
The following fluctuation has been introduced
for the metric tensor:
$$
g_{00}(\xb,t)=\one+\g(\xb,t)
\eqno(1)$$
with the Fourier expansion
$$
\g(\xb,t)=\sum_{\kb}\bigl(\ck e^{i(\kb\xb-\o t)}+c.c.\bigr),
\eqno(2)$$
where $\o =ck$. For the other components of the metric tensor
no definite suggestion was made; the knowledge of $g_{00}$ is
usually sufficient to describe nonrelativistic dynamics of masses.
For technical simplicity, we have set unity for the volume.
The complex coefficients $\{\ck\}$ are independent random variables
of zero mean. The stochastic averages of squared modules satisfy
the following relations:
$$
<\vert\ck\vert^{2}>=\cases {\Lambda^{4/3}k^{-5/3}, &$k<2\pi/\lc;$\cr
			    0,&$otherwise.$\cr}
\eqno(3)$$
where $\Lambda=\sqrt{G\hbar/c^3}\approx10^{-33}cm$ denotes the
Planck length and
$\lc$ is the cutoff parameter originally set to $10^{-12}-
10^{-13}cm$ [1-3]. Without a cutoff length the theory would be divergent.

In Refs.[1-3], the nonrelativistic Schr\"odinger equation
was considered on the random space-time (1). The effect of
$\g$ perturbing the metric tensor component $g_{00}$ is
equivalent to introducing the potential
$$
V(\xb,t)=\half Mc^2\matilde\g(\xb,t)
\eqno(4)$$
into the Schr\"odinger equation.
The tilde stands for averaging over the particle's volume.
According to the proposal of Refs.[1-3], the wave function
generally obeys the Schr\"odinger equation with the
potential (4) but, from time to time, instantaneous reduction
processes interrupt the ordinary dynamic evolution.

In the present paper we concentrate on the periods of dynamic
evolution between instantaneous reductions. We are going to
calculate the electromagnetic radiation which is due to the
electric charge of the particles, performing forced oscillation
influenced by the potential (4). We find that the radiation would be
surprisingly intensive and may  be moderate
only if the cutoff parameter $\lc$ is critically high. This result
may strengthen previous warnings [4,5] that Eqs.(1-3)
considerably overestimate the conceivable fluctuations
of the space-time metric. The effect calculated here seems to be direct
and inevitable consequence of the K\'arolyh\'azy model.
\bigskip
\line{\bf 2. Dipole radiation of oscillating charged particles\hfill}

In this paragraph we calculate the electromagnetic
radiation of a particle of charge $e$, performing oscillations
forced by the fluctuations of the hazy space-time (1).

We start from the dipole formula [6] for the radiation intensity:
$$
I_{\o}={4e^2\over3c^3}\vert\dot2\xbo\vert^2
\eqno(5)$$
where $\dot2\xbo$ is the Fourier transform of the acceleration
of the charged particle. The dipole approximation
is safe when the radiating charged source is much smaller
than the wavelength $\lambda$. In our considerations the sources
are the electrons and nuclei hence Eq.(5) remains valid well above
$\lambda\simeq10^{-13}cm$. (For atomic matter the electron shell as well as
the whole neutral structure has the
extension $\sim10^{-8}cm$. For greater wavelengths the system reacts as
globally neutral.)

It is known [6] that the dipole radiation (5) can equally
well be calculated from the classical acceleration of the
particle so, for the present purpose we shall
use the classical Newton
equation of motion instead of the Schr\"odinger one:
$$
\dot2\xb(t)=-{1\over M}\nabla V\bigl(\xb(t),t\bigr)+other\ forces.
\eqno(6)$$
To calculate  Fourier components of both sides,  we make
the following simplifying assumptions: i)
the amplitude of the forced oscillation is small
compared to the wavelength $\lambda$ of the driving field (4),
verified later, ii) other forces influencing the particle, as compared
to the gravitational driving force on RHS. of Eq.(6), are ignored.
Then, from Eqs.(2),(4) and (6), one obtains:
$$
\dot2\xbo=\half c^2(-i\kb)\ck e^{i\kb\xb}
\eqno(7)$$
Substituting this result into Eq.(5) and taking the
stochastic average according to Eq.(3) one gets
$$
<I_{\o}>={4e^2\over3c^3}\bigl<\vert\dot2\xbo\vert^2\bigr>=
{4e^2c\over3}\Lambda^{4/3}k^{1/3}.
\eqno(8)$$
There will be no radiation below the wavelength $\lc$ of the
spectrum of the driving force (4).

To calculate the spectral intensity of the  radiation, invoke
the well known rule:
$$
\sum_{\kb}\rightarrow4\pi\int d\lambda\lambda^{-4}.
\eqno(9)$$
So from Eq.(8) we obtain the final expression for the
spectral intensity of dipole radiation:
$${\it
\bigl<{dI\over d\lambda}\bigr>=\cases
{{16\over3}(2\pi)^{1/3}\pi e^2c\Lambda^{4/3}\lambda^{-13/3},
					    &$\lambda>\lc;$\cr
					  0,&$otherwise.$\cr}}
\eqno(10)$$
Consequently, the total intensity can be estimated as follows:
$$
<I>\ =\int_{\lc}^\infty dI\ \approx\  e^2c\Lambda^{4/3}\lc^{-10/3}
\eqno(11)$$
where a constant number factor of order unity is ignored.

We owe to justify assumptions i) and ii). From Eq.(7) we
obtain the Fourier transform of the particle's elongation:
$$
\xbo=\half c^2{i\kb\over \o^2}\ck e^{i\kb\xb}.
\eqno(12)$$
By using the above expression together with Eq.(3),
the range of the squared amplitude of the forced oscillation is:
$$
\bigl<\vert\D\xb\vert^2\bigr>\equiv
\sum_{\kb}
\bigl<\vert\xbo\vert^2\bigr>\approx
\Lambda^{4/3}\lc^{2/3}
\eqno(13)$$
Trying with the only reasonable cutoff values
$\lc=10^{-5}-10^{-13}cm$, the average oscillation amplitude will
be about $10^{-24}-10^{-26}cm$. This extremely small amplitude directly
justifies our assumption i). As for ii), when the particle is not free,
the forced oscillations, due to their extremely small amplitudes,
will simply be superposed onto the
nonrelativistic motion of particles. Up to this, extremely good,
approximation, the  dipole radiation of the forced oscillations will
not be affected by binding (or other) interactions. If in Eq.(6)
{\it other forces} act they will cause, e.g., thermal radiation
which will be incoherently superposed by the radiation (10).

Consequently, the
calculated radiation formula (10) itself can be extended to interacting
or even bound charged particles. Usually they will radiate decoherently,
each according to the Eq.(10), provided the wavelength $\lambda$ is
much smaller than the separation of the charged particles.
(An interesting exception is
the radiation of nuclei bound in ideal crystals where the driving
forces are strongly correlated even at wavelengths much smaller
than the lattice constant.)
\bigskip
\line{\bf 3. Discussion \hfill}

For calculating the outcoming radiation, one has to multiply the
intensity (11) with the density of charges present, and integrate up for the
volume of the source. We count only the charges which are free or bound in a
system bigger than $\lc$. First consider the range
$10^{-13}cm\le\lc\le10^{-8}cm$.
Then all charged particles of the atomic and even condensed matter would
contribute to radiation decoherently since the average separation
of charges (electrons, nuclei) is bigger than $\lc$. According to Eq.(11),
some $10^{23}$ charged particles of a mole (e.g. several grams) of any
condensed matter would produce a radiation with an overall intensity
 $\sim10^{10}erg/s$ if $\lc=10^{-12}cm$ or, still a considerable value
$\sim1erg/s$ if $\lc=10^{-9}cm$. In the spectrum the shortwave end
$\lambda\approx\lc$ would dominate, so this radiation would mean hard
$\gamma$ or X-rays: $\sim10^{15} \gamma$-photons if $\lc=10^{-12}cm$ or
$\sim10^8$ R\"ontgen-photons if $\lc=10^{-9}cm$, per each mole.

Such
number of hard photons is a dangerous radiation from e.g. lead used for
shielding against $\gamma$-ray radiation, which would have been discovered
long ago. Therefore certainly
$$
\lc\ >\ 10^{-8}cm.
$$

A great number of charged particles separated at larger than the above
distance
can be most typically found in plasmas.
Consider a gas at 1 atmosphere, heated up above $3000K$.
Then it is in ionized plasma state and the average charge separation
is cca. $10^{-6}cm$. Then for $\lc\approx10^{-6}cm$ from one mole
(e.g. cca. $1/4m^3$) of hot gas the radiation would be $10^{-10}erg/s$,
i.e. $\sim10$ photons per seconds, each of energy $\sim100eV$.

This intensity is low; however at $3000K$ the peak of the plasma's
thermal radiation is in the near infrared, while the $100eV$ photons
are somewhere between UV and X-rays. Their calculated intensity
is by some $100$ (!) orders of magnitude higher than the intensity of
the thermal photons of the same wavelength.

Such nonthermal hard UV radiation should have been picked up
by detectors long ago, and it has not been. So
$$
\lc\ >\ 10^{-6}cm.
$$
{}From the very essence of the model of K\'arolyh\'azy $et\ al.$ [1-3]
follows that the cutoff parameter ${\lc}$ should not be macroscopic,
see, e.g., in Ref.[7], too.
Hence the remaining range is, e.g., $10^{-6}cm<\lc<10^{-5}cm$. Here
it seems that the inevitable electromagnetic radiation would not
necessarily result in trivially drastic effects. (Namely, for the plasma
experiment, one mole of dilute plasma with $10^{-5}cm$ average
separation of ions would occupy a container of cca. $250m^3$ while
the photon flux would be cca. one photon of $10\ eV$ in a minute.)

So the fact that drastic UV radiation from very familiar kinds of matter
around us and in laboratories is generally not detected leaves
for the cutoff length
necessary in the K\'arolyh\'azy model [1-3]  a narrow
range $$10^{-6}cm<\lc<10^{-5}cm$$. Unfortunately this range seems to be
excluded by cosmological considerations listed in a previous
paper [5] of ours.
\bigskip
The necessity and timeliness to perform the present research
were recognized in a discussion with Prof. P. Gn\"adig of the
E\"otv\"os University. This work was supported by the
Hungarian Scientific Research Fund under Grant OTKA 1822/1991.
\bigskip
\line{\bf References\hfill}
\frenchspacing
\item{[1]} F.K\'arolyh\'azy, NuovoCim. {\bf XLIIA}, 1506 (1966)\smallskip
\item{[2]} F.K\'arolyh\'azy,A.Frenkel and B.Luk\'acs, in:Physics as
natural philosophy, eds. A.Shi\-mony and H.Feschbach (MIT Press,
Cambridge,MA,1982)\smallskip
\item{[3]} F.K\'arolyh\'azy,A.Frenkel and B.Luk\'acs, in:Quantum
concepts in space and time, eds. R.Penrose and C.J.Isham
(Clarendon,Oxford,1986)\smallskip
\item{[4]} L.Di\'osi and B.Luk\'acs, Phys.Lett. {\bf 142A},331(1989)
	   \smallskip
\item{[5]} L.Di\'osi and B.Luk\'acs, preprint KFKI-1993-05/AB
\item{[6]} L.Landau and E.Lifchitz,Th\'eorie des Champs (Mir,Moscou,1970)
	   \smallskip
\item{[7]} G.C.Ghirardi,A.Rimini and T.Weber, Phys.Rev.{\bf D34},
470 (1986)\smallskip
\vfill
\end